\documentclass[vecphys]{svmult}

\usepackage{makeidx}         % allows index generation
\usepackage{graphicx}        % standard LaTeX graphics tool
\usepackage{multicol}        % used for the two-column index
\usepackage[bottom]{footmisc}% places footnotes at page bottom
\usepackage{latexsym}
\usepackage{dsfont}
\usepackage{overpic}

\makeindex             % used for the subject index
\newcommand{\beq}{\begin{equation}}
\newcommand{\eeq}{\end{equation}}
\newcommand{\beqa}{\begin{eqnarray}}
\newcommand{\eeqa}{\end{eqnarray}}
\newcommand{\no}{\nonumber}

\newcommand{\Lagr}{\mathcal{L}}
\newcommand{\sfrac}[2]{{\textstyle\frac{#1}{#2}}}
\newcommand{\MeV}{\,\mathrm{MeV}}
\newcommand{\GeV}{\,\mathrm{GeV}}

\begin{document}

\title*{Introduction to Chiral Perturbation Theory}
\author{Bu{\=g}ra Borasoy}
\institute{Helmholtz-Institut f\"ur Strahlen- und Kernphysik, Universit\"at Bonn,\\
Nussallee 14-16, D-53115 Bonn, Germany; \
\texttt{borasoy@itkp.uni-bonn.de}}
\maketitle

A brief introduction to chiral perturbation theory, the effective field theory
of quantum chromodynamics at low energies, is given.

%%%%%%%%%%%%%%%%%%%%%%%%%%%%%%%%%%%%%%%%%%%%%%%%%%%%%%%%%%%%%%%%%%%%%%%%%%%
\section{Introduction}
\label{sec:intro}

The strong interactions are described by 
quantum chromodynamics (QCD), a local non-abelian gauge theory.
The QCD Lagrangian comprises quark and gluon fields which
carry {\it color} charges and interact with coupling strength $g$.
The renormalized coupling $g$ depends on the momentum at which the measurement is performed
and decreases as the momentum scale $Q$ is increased.
This behavior is referred to as {\it running}   of the strong coupling constant 
$\alpha_s (Q) = g^2(Q)/(4 \pi)$.
The coupling $\alpha_s$ decreases for large momenta
and the theory becomes {\it asymptotically free} with  quasi-free quarks and gluons \cite{asympfree}.

\begin{figure}[ht]
\centering
\parbox{4cm}{\centering\includegraphics[width=1.8cm]{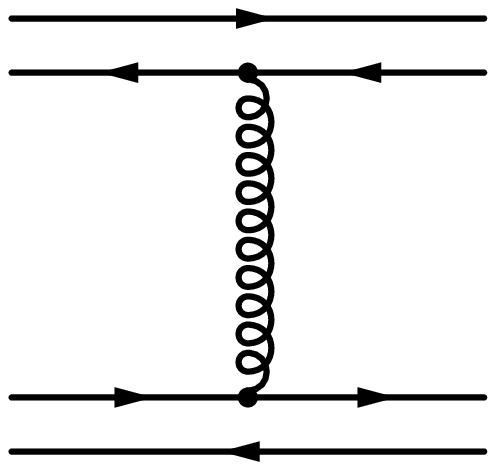}}    
  \parbox{3cm}{\centering\includegraphics[width=1.5cm]{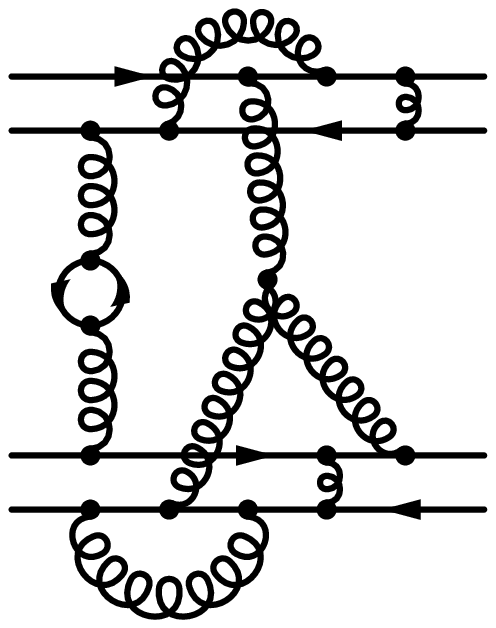}}
\caption{Two sample diagrams which contribute to $\pi \pi$ scattering.
         Solid and curly lines denote quarks and gluons, respectively.}
\label{fig:pipiscatt}
\end{figure}

In this regime of QCD perturbation theory in $\alpha_s$ converges.
For small momenta, on the other hand, $\alpha_s$ is large so that quarks and gluons 
arrange themselves in strongly bound clusters to form hadrons, e.g.,
protons, neutrons, pions, kaons, etc.
In order to describe the physics of hadrons at low energies,
perturbation theory is not useful because $\alpha_s$ is large.
This is illustrated for $\pi \pi$ scattering in Fig.~\ref{fig:pipiscatt}
where both sample diagrams---along with infinitely many other contributions---are 
equally important, although the right diagram appears at a
much higher order in the perturbative series in $\alpha_s$.

Alternative model-independent approaches are required
in the non-pertur-bative regime of QCD.
These are provided either by QCD lattice simulations which are a numerical solution to QCD
or---at low energies---by chiral perturbation theory, the effective field theory of QCD.
In the first case the QCD path integral in Euclidean space-time is evaluated numerically
via Monte Carlo sampling, see e.g. \cite{lattice}.
In the latter case, one makes use of the fact that
at low energies the relevant, effective degrees of freedom are hadrons
rather than quarks and gluons which are not observed as free particles.

It is thus convenient to replace in the low-energy limit the QCD Lagrangian by an effective Lagrangian
which is formulated in terms of the effective degrees of freedom, i.e. 
pions, kaons, eta, etc.
The corresponding field theoretical formalism is called chiral perturbation theory (ChPT)
\cite{Wein1, GL1, GL2}.

In these lectures a brief introduction to ChPT is presented emphasizing
some basic principles and a few simple applications. It is not intended
to provide a detailed review of ChPT,
in particular we restrict ourselves to the purely mesonic sector and do not 
consider baryons. For more comprehensive reviews
the reader is referred to \cite{ChPT}.

This paper is organized as follows. In the next section some well-known examples
and basic principles of effective field theories in general are presented.
Section~\ref{sec:ConChiEL} describes the construction principles for the
chiral effective Lagrangian. Higher orders and loops are discussed in
Sec.~\ref{sec:HiOrdLoop}.

%%%%%%%%%%%%%%%%%%%%%%%%%%%%%%%%%%%%%%%%%%%%%%%%%%%%%%%%%%%%%%%%%%%%%%%%%%%
\section{Effective field theories}
\label{sec:EFT}

The basic idea of an effective field theory is to treat the active, light
particles as relevant degrees of freedom, while the heavy particles are
frozen and reduced to static sources.
The dynamics are described by an effective Lagrangian which
is formulated in terms of the light particles and incorporates all important
symmetries and symmetry-breaking patterns of the underlying fundamental theory.

\subsection{Scattering of light by light in QED at very low energies}
\label{subsec:EFTQED}

The Lagrangian of quantum electrodynamics (QED) is given by
\beq
\Lagr_{\mbox{\scriptsize{QED}}} = \Lagr_0 + \Lagr_{\mbox{\scriptsize{int}}}
\eeq
with the free part
\beq
\Lagr_0 = \bar{\psi} \left( i \partial \!\!\! /  -m  \right) \psi - \frac{1}{4} F_{\mu \nu} F^{\mu \nu}
\eeq
and the interaction piece
\beq
\Lagr_{\mbox{\scriptsize{int}}} = -e \bar{\psi}  A \!\!\! / \psi \ .
\eeq
Fermion and photon fields are denoted by 
$\psi$ and $A_\mu$ , respectively, 
$F_{\mu \nu} = \partial_\mu A_\nu - \partial_\nu A_\mu$ is the field strength tensor, and 
a gauge fixing term has been omitted for brevity.

\begin{figure}[ht]
\centering
\includegraphics[width=2cm]{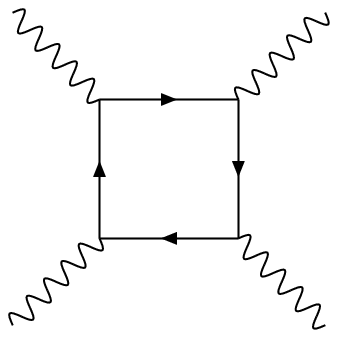}
\caption{Light by light scattering to lowest order. The wavy and solid lines denote the 
photons and electrons, respectively.}
\label{fig:lightscat}
\end{figure}

Consider light by light scattering at very low photon energies $\omega \ll m$.
In this instance, electrons (and positrons) cannot be produced in the final
state, but contribute instead via virtual processes.  
The calculation of the lowest order
diagram which is given by a single electron loop, Fig.~\ref{fig:lightscat},
is straightforward but cumbersome.
However, at very low energies the amplitude for light by light 
scattering is equally reproduced by the effective Lagrangian \cite{HE,Schw}
\beq  \label{eq:EHlagr}
\Lagr_{\mbox{\scriptsize{eff}}} = - \frac{1}{4} F_{\mu \nu} F^{\mu \nu}
           + \frac{e^4}{1440 \, \pi^2 m^4} \left[ ( F_{\mu \nu} F^{\mu \nu} )^2 
           + \frac{7}{16} ( F_{\mu \nu} \tilde{F}^{\mu \nu})^2  \right]  + \ldots 
\eeq
which only contains the field strength tensor $F_{\mu \nu}$ and its
dual counterpart $\tilde{F}_{\mu \nu} = \epsilon_{\mu \nu \rho \sigma} F^{\rho \sigma}$
as explicit degress of freedom. The ellipsis denotes corrections to this 
Lagrangian involving more derivatives which arise from the energy expansion
of the original one-loop diagram in powers of $\omega/m$. 
Moreover, the coefficients of the operators in the effective Lagrangian
receive corrections of higher orders in $e^2$ through multiloop diagrams.

\begin{figure}[ht]
\centering
\parbox{4cm}{\centering\includegraphics[width=1.8cm]{lightscat.eps}}   $\rightarrow$ 
\parbox{3cm}{\centering\includegraphics[width=1.5cm]{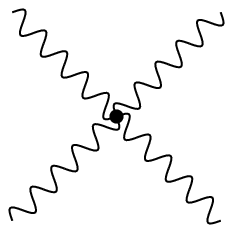}}
\caption{The one-loop diagram of QED is replaced in the effective theory
         by 4-photon contact interactions.}
\label{fig:lightcont}
\end{figure}

It is instructive to illustrate the conversion to the effective field theory
with Feynman diagrams.
By treating at very low photon energies the electrons as heavy static sources
the electron propagators of the electron loop in QED ``shrink'' to a single point.
This gives rise to 4-photon contact interactions
which correspond to the vertices of the effective Lagrangian, see Fig.~\ref{fig:lightcont}.

A significant property is that the U(1) gauge symmetry of the underlying
QED Lagrangian is maintained by the effective Lagrangian, Eq.~(\ref{eq:EHlagr}), since the building blocks
$F_{\mu \nu} F^{\mu \nu}$ and $ F_{\mu \nu} \tilde{F}^{\mu \nu}$ are both
gauge invariant. As we will see below, invariance under the relevant symmetries
is an important constraint in constructing effective Lagrangians.

\subsection{Weak interactions at very low energies}
\label{subsec:EFTWeak}

\begin{figure}[ht]
\centering
\parbox{4cm}{\centering\includegraphics[width=2.cm]{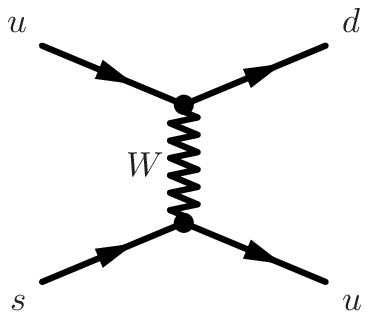}}   $\rightarrow$ 
  \parbox{3cm}{\centering\includegraphics[width=2.cm]{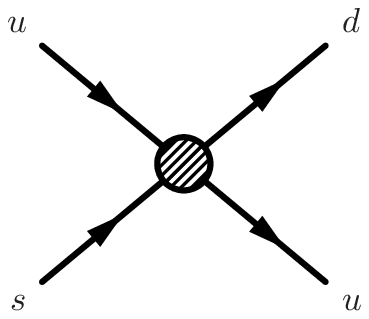}}
\caption{At low energies the single $W$ boson exchange reduces to a four-quark contact interaction.}
\label{fig:Wexch}
\end{figure}

A second well-known example of an effective field theory is encountered in
weak interactions. Consider the amplitude for the flavor changing weak process at lowest order from single 
$W$ boson exchange
\beq
{\cal A} = \left(\frac{i g}{\sqrt{2}}\right)^2 V_{us} V_{ud}^* \left(\bar{u} \gamma^\mu 
              \sfrac{1 - \gamma_5}{2} s \right) \left(\bar{d} \gamma^\nu 
              \sfrac{1 - \gamma_5}{2} u \right) \left( \frac{- i g_{\mu \nu}}{p^2 - M_W^2} \right) \ ,
\eeq
where $V_{ij}$ are elements of the Kobayashi-Maskawa mixing matrix
and the $W$ propagator is given in Feynman gauge.
In the limit of small momentum transfer, $p^2 \ll M_W^2$, the $W$ propagator
can be expanded in $p^2/M_W^2$ such that the amplitude is approximated by the local
interaction
\beq
{\cal A} = \frac{i}{M_W^2} \left(\frac{i g}{\sqrt{2}}\right)^2 V_{us} V_{ud}^* \left(\bar{u} \gamma^\mu 
              \sfrac{1 - \gamma_5}{2} s \right) \left(\bar{d} \gamma_\mu 
              \sfrac{1 - \gamma_5}{2} u \right) + {\cal O} \left( \frac{p^2}{M_W^4} \right) \ .
\eeq
Diagrammatically this approximation is illustrated in Fig.~\ref{fig:Wexch},
where the contact interaction arises from the effective Lagrangian
\beq
\Lagr_{\mbox{\scriptsize{eff}}} = -2\sqrt{2} G_F V_{us} V_{ud}^* \left(\bar{u} \gamma^\mu 
              \sfrac{1 - \gamma_5}{2} s \right) \left(\bar{d} \gamma_\mu 
              \sfrac{1 - \gamma_5}{2} u \right)
\eeq
with the Fermi constant $G_F = g^2/\left( 4 \sqrt{2} M_W^2 \right)$.

\subsection{Chiral symmetry in QCD}
\label{subsec:ChiSymQCD}

As mentioned above, the relevant symmetries of the underlying theory 
must also be maintained by the effective field theory.
In this section, we will study the (approximate) chiral symmetry of QCD.
The QCD Lagrangian reads in compact notation
\beq
\Lagr_{\mbox{\scriptsize{QCD}}} = \bar{q} \left( i \gamma^\mu D_\mu - m_q \right) q
  - \frac{1}{2} \mbox{Tr} _c \left( G^{\mu \nu} G_{\mu \nu} \right) \ ,
\eeq
where $q^T = (u,d,s,c,b,t)$ comprises the six quark flavors, 
$D_\mu = \partial_\mu - i g G_\mu$ is the covariant derivative,
$G_\mu$ the gluon fields, and 
$G_{\mu\nu} = \partial_\mu G_\nu - \partial_\nu G_\mu  
- i g \, [G_\mu , G_\nu]$ the gluon field strength tensor.
Tr$_c$ denotes the trace in color space.
The Dirac field $q$ is a 72-component object; 
each of the 6 quark flavors appears in 3 different colors and has 4 spinor components.

The quarks can be grouped into light and heavy flavors according to their masses:
the $u,d,s$ quarks are substantially lighter than the $c,b,t$ quarks \cite{pdg}.
Hence, the limit of massless light quarks, $m_u= m_d= m_s=0$, the so-called chiral limit,
seems to be a reasonable approximation and can be improved by treating the light quark masses
as perturbations.
The $c,b,t$ quarks, on the other hand, can be treated at low energies as 
infinitely heavy and the only active degrees of freedom are those associated
with the light $u,d,s$ quarks.

It is straightforward to see that in the chiral limit the QCD Lagrangian has an 
extra symmetry. 
In this limit, the relevant part of $\Lagr_{\mbox{\scriptsize{QCD}}}$ is
(we use the same notation for simplicity)
\beq
\Lagr_{\mbox{\scriptsize{QCD}}}= \sum_{q=u,d,s} \bar{q} i \gamma_\mu D^\mu  q
        - \frac{1}{2} \mbox{Tr}_c ( G_{\mu \nu} G^{\mu \nu}) \ .
\eeq
Here, $q$ represents a one-flavor quark field.
By introducing right- and left-handed quark fields
\beq
q_{R/L} = \, \frac{1}{2} ( 1\pm \gamma_5) \, q  
\eeq
one arrives at
\beq
\Lagr_{\mbox{\scriptsize{QCD}}}= \sum_{q=u,d,s} \left( \bar{q}_L i \gamma_\mu D^\mu  q_L
       +\bar{q}_R i \gamma_\mu D^\mu  q_R \right)
        - \frac{1}{2} \mbox{Tr}_c ( G_{\mu \nu} G^{\mu \nu}) \ .
\eeq
Independent transformations of the right- and left-handed quark fields
\beq
q_R \, \rightarrow  \, R \, q_R \ , \qquad \qquad 
q_L \, \rightarrow  \, L \, q_L
\eeq
with $R \in \mbox{SU(3)}_R, \ L \in \mbox{SU(3)}_L$ leave the massless QCD Lagrangian invariant.
This invariance is referred to as 
SU(3)$ _L \times $SU(3)$_R$ chiral symmetry of massless QCD.
One observes that the gluon interactions do not change the helicity of
quarks but the quark mass term does.

Due to Noether's theorem an immediate consequence of a continuous symmetry of a Lagrangian is
the existence of a conserved current $J_\mu$ with $\partial_\mu J^\mu =0$.
The corresponding charge 
\beq
Q(t) = \int d^3 x \ J_0(t, \mathbf{x})
\eeq
is time-independent, i.e. $d Q / dt=0$.
Familiar examples are the invariance of the Lagrangian with regard
to translations in time and space and rotations which imply, respectively,
conservation of energy, momentum and angular momentum.
At the operator level, the conserved charges commute with the Hamiltonian.

In the chiral limit of QCD the conserved currents of chiral symmetry are
\beq
L_\mu^a = \sum_{q=u,d,s}  \bar{q}_L  \gamma_\mu \frac{\lambda^a}{2}  q_L \ ,
             \qquad  \quad R_\mu^a = \sum_{q=u,d,s}  \bar{q}_R  \gamma_\mu \frac{\lambda^a}{2}  q_R 
\eeq
with the Gell-Mann matrices $\lambda^a$.
The invariant charges $Q_L^a, Q_R^a$ generate the algebra of
SU(3)$_L$ and SU(3)$_R$, respectively.
It is useful to define the combinations
\beq
Q_V^a  =  Q_R^a + Q_L^a \ ; \qquad
Q_A^a  =  Q_R^a - Q_L^a
\eeq
which have a different behavior under parity
\beq
Q_V^a  \rightarrow  Q_V^a \ ; \qquad
Q_A^a  \rightarrow  - Q_A^a  \ .
\eeq
Consider an eigenstate $| \psi \rangle $ of $H_{\mbox{\scriptsize{QCD}}}$ (in the chiral limit)
\beq
  H_{\mbox{\scriptsize{QCD}}} | \psi \rangle = E | \psi \rangle \ .
\eeq
The states  $Q_V^a | \psi \rangle $ and  $Q_A^a | \psi \rangle $
have the same energy $E$ but opposite parity.
Thus for each positive parity state there should be
a negative parity state with equal mass. This pattern is, however, not observed in the particle spectrum
\cite{pdg}.
For example, the light pseudoscalar ($J^P = 0^-$) mesons, $(\pi , K, \eta)$,
have a considerably lower mass than the scalar ($J^P = 0^+$) mesons.% 

The solution to this paradoxon is provided by the Nambu-Goldstone realization of chiral symmetry \cite{NJL}
which asserts that the QCD vacuum, $|0\rangle$, is not invariant under the action
of the axial charges
\beq
Q_V^a |0\rangle = 0 \qquad  Q_A^a |0\rangle \ne 0 \ .
\eeq
The chiral $SU(3)_L \times SU(3)_R$ symmetry of the QCD Hamiltonian
is said to be {\it spontaneously} broken down to $SU(3)_V$.
Spontaneous breakdown of a symmetry takes place if the full symmetry group of the Hamiltonian
is not shared by the vacuum.

Another example of spontaneous symmetry breakdown occurs in ferromagnets.
For temperatures above the Curie temperature, $T > T_c$, the magnetic dipoles are
randomly oriented. As soon as the temperature falls below the Curie temperature $T_c$
spontaneous magnetization occurs and the dipoles are aligned in some 
arbitrary direction. Spontaneous symmetry breakdown takes also place for the 
SU(2)$_L \times$U(1) symmetry of the electroweak interactions.

In general, spontaneous breakdown of a continuous symmetry has important consequences.
Goldstone's theorem states that a spontaneously broken continuous symmetry implies
massless spinless particles: the Goldstone bosons.
In the case of massless QCD, the eight axial charges $Q_A^a$ create states 
$| \phi \rangle = Q_A |0\rangle $ which are energetically degenerate with the
vacuum $|0\rangle$ since 
\beq
 H | \phi \rangle = H Q_A |0\rangle = Q_A H |0\rangle = 0 \ .
\eeq
This gives rise to eight massless pseudoscalar mesons.
The axial charges $Q_A^a$ acting on any particle state generate Goldstone bosons, e.g. 
an energy eigenstate $| \psi \rangle$ is degenerate with the
multi-particle state $Q_A^a | \psi \rangle $ which resolves the paradoxon from above.

The eight lightest hadrons are indeed the pseudoscalars $\pi^\pm$,$\pi^0$,$K^\pm$,$K^0$,$\bar{K}^0$,\\
$\eta$
with masses $m_\pi \approx 138$ MeV, $m_K \approx 495$ MeV and $m_\eta \approx 547$ MeV \cite{pdg}.
Since the nonzero masses of the light quarks break chiral symmetry explicitly
the Goldstone bosons are not exactly massless.
However, the explicit breaking can be considered to be small and treated
perturbatively. In the limit of vanishing quark masses,
$m_u, m_d, m_s \rightarrow 0$, the Goldstone boson masses approach zero,
$m_\pi, m_K, m_\eta \rightarrow 0$, while all other hadrons remain massive in the chiral limit and 
are separated from the ground state roughly by a characteristic gap
\beq
\Delta \sim M_{\mbox{\scriptsize proton}} \sim 1 \, \mbox{GeV} \ .
\eeq

In the remainder of this section, it is demonstrated that only the spontaneous
breakdown of a {\it continuous} symmetry gives rise to Goldstone bosons, 
whereas in the case of a discrete symmetry Goldstone bosons are not generated.\\

\begin{figure}[ht]
\centering
\begin{tabular}{cp{1.8cm}c}
\begin{overpic}[width=0.2\textwidth,clip]{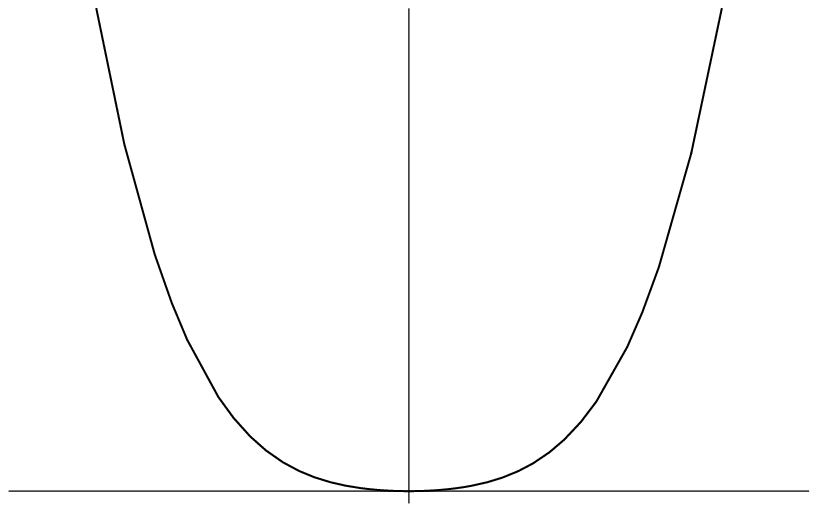}
  \put(90,-10){\scalebox{0.9}{$\phi$}}
  \put(52,58){\scalebox{0.9}{$V(\phi^2)$}}
\end{overpic} & &
\begin{overpic}[width=0.2\textwidth,clip]{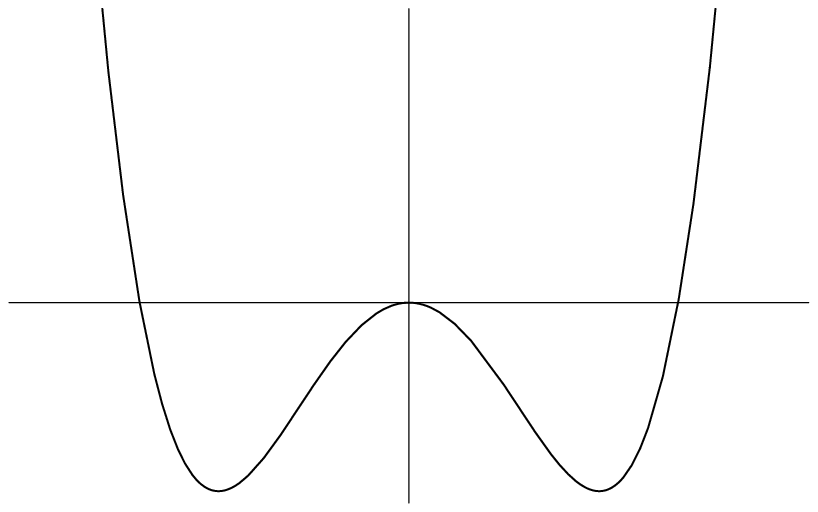}
  \put(92,13){\scalebox{0.9}{$\phi$}}
  \put(52,58){\scalebox{0.9}{$V(\phi^2)$}}
\end{overpic} \\[0.1cm]
\end{tabular}
\caption{Potential $V(\phi^2)$, Eq.~(\ref{eq:potclass}), for $m^2<0$ (left) and $m^2>0$ (right).}
\label{fig:break}
\end{figure}

\paragraph{Discrete symmetry case}
Consider the Lagrangian density with a scalar field $\phi$
\beq
\Lagr= \frac{1}{2} \partial_\mu \phi \partial^\mu \phi  
+ \frac{1}{2} m^2 \phi^2 - \frac{1}{4} \lambda \phi^4 \ .
\eeq
The Lagrangian is invariant under the discrete symmetry of reflections, $\phi \rightarrow - \phi$.
The corresponding potential is given by
\beq  \label{eq:potclass}
V(\phi^2) = - \frac{1}{2} m^2 \phi^2 + \frac{1}{4} \lambda \phi^4 \ ,
\eeq
and since the energy must be bound from below the coupling $\lambda$ is positive.
The coefficient $m^2$, on the other hand, is not constrained. There are two possible cases
depending on the sign of $m^2$ as illustrated in Fig.~\ref{fig:break}.
For $m^2<0$ there is a unique minimum at $\phi=0$, but for $m^2>0$ the potential $V(\phi^2)$
is minimized by two possible ground state fields $\phi = \pm \sqrt{m^2/ \lambda}$.
In the quantum field theoretical language this implies that the field $\phi$
develops a vacuum expectation value
\beq
\langle 0|\phi | 0 \rangle = \pm \sqrt{\frac{m^2}{\lambda}} \ .
\eeq
Hence, there are two possible vacua but each vacuum is not invariant under
reflection symmetry, i.e. the theory is spontaneously broken. Massless Goldstone
bosons, however, do not appear.\\

\paragraph{Continuous symmetry case}
Consider now the Lagrangian with two scalar fields $\sigma$ and $\pi$
\beq
\Lagr= \frac{1}{2} (\partial_\mu \sigma)^2 + \frac{1}{2}  (\partial_\mu \pi)^2
        - V(\sigma^2 + \pi^2) 
\eeq
with $V$ defined as in Eq.~(\ref{eq:potclass}).
It exhibits an O(2) symmetry; continuous transformations of the type
\beq
\left(\begin{array}{c}  \sigma\\ \pi\end{array}\right) \rightarrow
  \left(\begin{array}{cc} \phantom{+} \cos \alpha  &\phantom{+} \sin \alpha \\
  -\sin \alpha &  \phantom{+} \cos \alpha  \end{array}\right)
  \left(\begin{array}{c}  \sigma\\ \pi\end{array}\right)
\eeq
leave the Lagrangian invariant.
The extrema of the corresponding potential $V$ are determined  by the equations
\beqa
  \frac{d V}{d \sigma} &=& \sigma [ -m^2 + \lambda (\sigma^2 + \pi^2) ] =0 \ , \no \\
  \frac{d V}{d \pi} &=& \pi [ -m^2 + \lambda (\sigma^2 + \pi^2) ] =0 \ .
\eeqa
For $m^2 > 0$ the minima are at $ \sigma^2 + \pi^2 =  m^2/ \lambda  $
and related to each other through O(2) rotations.
Any point on the circle of minima may be chosen to be the true vacuum $|0 \rangle$.
One may take, e.g.,
\beq
\langle 0|\sigma | 0 \rangle =  \sqrt{\frac{m^2}{\lambda}} \ ; \qquad
   \langle 0|\pi | 0 \rangle = 0 \ .
\eeq
Clearly,  the O(2) symmetry of the Lagrangian is spontaneously broken by the vacuum state.
Small oscillations around this vacuum state can be described by shifting the
$\sigma$ field
\beq
\sigma' \equiv  \sigma - \sqrt{\frac{m^2}{\lambda}} 
\eeq
so that the Lagragian reads in terms of the new fields (up to an irrelevant constant)
\beqa
\Lagr  &=& \, \frac{1}{2} (\partial_\mu \sigma')^2 + \frac{1}{2}  
               (\partial_\mu \pi)^2 - m^2 \sigma'^2 \no \\
  && -\lambda  \, \sqrt{\frac{m^2}{\lambda}} \, \sigma' \, (\sigma'^2 + \pi^2) 
  -  \frac{1}{4} \,  \lambda \, (\sigma'^2 + \pi^2)^2 \ .
\eeqa
With this choice of coordinates the mass term for the $\pi$ field has disappeared
and the $\pi$ becomes massless. The $\pi$ field is then interpreted as a 
polar angle oscillation around the vacuum which does not cost any energy.
A Goldstone boson has been created through spontaneous breakdown of the
continuous O(2) symmetry in the original Lagrangian.

%%%%%%%%%%%%%%%%%%%%%%%%%%%%%%%%%%%%%%%%%%%%%%%%%%%%%%%%%%%%%%%%%%%%%%%%%%%
\section{Construction of the chiral effective Lagrangian}
\label{sec:ConChiEL}

In this section we outline the construction principles for the chiral effective
Lagrangian. The chiral SU(3) Lagrangian is in general a function of the Goldstone boson (GB) fields 
$(\pi^0, \pi^\pm, K^\pm, K^0, \bar{K}^0, \eta)$. In order to construct 
the effective Lagrangian, we must first know the interaction between the GBs.

To this aim, we recall from the previous section that the eight 
axial charges $Q_A^a$ do not annihilate the vacuum, $Q_A^a | 0 \rangle \ne 0$.
The states $Q_A^a | 0 \rangle \ne 0$ are associated with the GBs 
$\phi^a=(\pi, K,\eta)$.
This implies non-vanishing matrix elements of the axial vector current $A_\mu^a$
\beq  \label{eq:MEAVC}
\langle 0 | A_\mu^a (x) | \phi^b (p) \rangle = i e^{- i p \cdot x} \, p_\mu \, \delta^{ab} \, f_a 
\eeq
(no summation over $a$).
The decay constant $f_a$ measures the strength with which the 
Goldstone boson $\phi^a$ decays via the axial vector current $A_\mu^a$ into the hadronic vacuum. 
The decay constants are extracted experimentally from weak decays of the GBs,
e.g., $\pi^+ \to l^+ \nu_l$ yields $f_\pi = 92.4$ MeV \cite{Hol}.

\begin{figure}[ht]
\centering
  \parbox{4cm}{\centering \includegraphics[width=2cm]{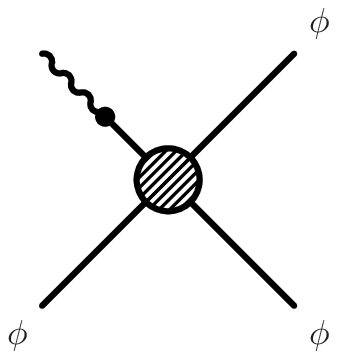}} \qquad
  \parbox{4cm}{\centering \includegraphics[width=2cm]{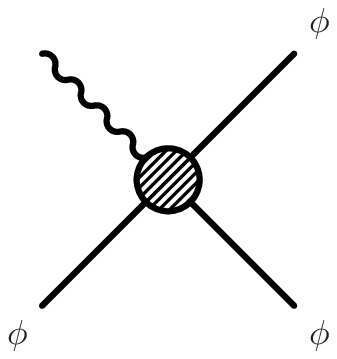}}
\caption{Contributions to ${\cal M}_\mu$ in Eq.~(\ref{eq:avphi3}) with (left) and without (right)
         a GB pole. The wavy and solid lines denote the axial vector current and the Goldstone bosons,
         respectively.}
\label{fig:avphi3}
\end{figure}

Taking the divergence of Eq.~(\ref{eq:MEAVC}) leads to
\beq   \label{eq:divME}
\langle 0 | \partial^\mu A_\mu^a (0) | \phi^b (p) \rangle =  \delta^{ab} m_a^2 f_a \ .
\eeq
In the chiral limit, the axial vector current is conserved, $\partial^\mu A_\mu^a = 0$,
so that $m_a^2 = 0$ as required by Goldstone's theorem.
In the real world, however, chiral SU(3)$_L \times$ SU(3)$_R$ symmetry is explicitly broken 
by the finite quark masses $m_u,m_d,m_s$ and the axial vector current is not conserved.
One introduces the GB field operators $\Phi^a$ with the normalization  
$\langle 0 |\Phi^a (0) | \phi^b (p)\rangle = \delta^{ab}$. Eq.~(\ref{eq:divME}) can then be
rewritten as
\beq
\langle 0 | \partial^\mu A_\mu^a (0) | \phi^b (p) \rangle =  
  m_a^2 f_a \langle 0 |\Phi^a (0) | \phi^b (p)\rangle  \  .
\eeq
At the operator level, this is the hypothesis of the partially conserved axial vector current (PCAC)
\beq   \label{eq:PCAC}
\partial^\mu A_\mu^a =  m_a^2 f_a \Phi^a \ .
\eeq
The axial currents can thus be employed as interpolating fields for the
Goldstone bosons and identity (\ref{eq:PCAC}) implies a vanishing
interaction between the GBs at zero momentum.
Consider to this end, e.g., the matrix element (suppressing flavor indices)
\beq  \label{eq:avphi3}
{\cal M}_\mu (p_1,p_2,p_3) = \langle \phi (p_2) \phi (p_3) | A_\mu (0) | \phi (p_1) \rangle \ .
\eeq
The amplitude ${\cal M}_\mu$ contains two parts: contributions with no GB poles
and contributions where the axial current generates a GB pole, see Fig.~\ref{fig:avphi3}.
In the chiral limit, the matrix element has the decomposition
\beq
{\cal M}_\mu (p_1,p_2,p_3) = \frac{f q_\mu}{q^2} T (p_1,p_2,p_3,q) + R_\mu \ ,
\eeq
where $q = - p_1 - p_2 - p_3$, $T$ is the GB-GB scattering matrix element,
$f$ the decay constant in the chiral limit, and
$R_\mu$ is non-singular as $q_\mu \to 0$ by definition.
Contracting both sides with $q^\mu$ yields
\beq
0 = q^\mu {\cal M}_\mu (p_1,p_2,p_3) = f T (p_1,p_2,p_3,q) + q^\mu R_\mu \ .
\eeq
In the limit $q_\mu \to 0$ one obtains
\beq
T (p_1,p_2,p_3,q) = 0 \ .
\eeq
The GBs do not interact at vanishing momenta.

At low but finite energies, the interaction between GBs can be expanded in powers of small
momenta. Consider for example the GB-GB scattering matrix $T$ which can be written as a function
of the three Mandelstam variables $s = (p_1+p_2)^2$, $t = (p_1+p_3)^2$ and $u = (p_1+q)^2$. 
Its low energy expansion reads
\beq
T(s,t,u) = f_1 s + g_1 t + h_1 u + \ldots 
\eeq
with momentum-independent expansion coefficients $f_i, g_i, h_i$.
The chiral effective Lagrangian is also ordered according to the low energy expansion.
Powers of GB momenta in the amplitude correspond to powers of derivatives on GB fields
in the Lagrangian.
The ordering of the effective Lagrangian in increasing powers of derivatives
is called {\it chiral ordering} or {\it chiral power counting}.

Next, we would like to investigate how GB fields are represented in the chiral Lagrangian.
To this aim, we shall study the transformation properties of the GBs under chiral transformations.

Let $G$ be the group of chiral SU(3)$_L \times$ SU(3)$_R$ transformations.
For a given representation of $G$ the GB fields transform according to
\beq
\phi \to \phi' = F (g, \phi) \ , \qquad     g \in G
\eeq
with the representation property
\beq
F (g_1, F(g_2, \phi)) = F (g_1 g_2, \phi) \ .
\eeq
Consider group elements $h \in G$ which leave the ``origin'', i.e. the vacuum,
invariant, $F (h, 0) = 0$. Obviously, these elements form a subgroup $H$: for
$h_1, h_2 \in H$ it follows that $h_1 h_2 \in H$. 
$H$ is equivalent to the subgroup SU(3)$_V$ which leaves the vacuum invariant.

The function
\beq
g \to F(g,0) = F(gh,0)   \qquad h \in H 
\eeq
maps the coset space $G/H$ onto the space of GB fields.
This mapping is invertible since
$ F(g_1,0) = F(g_2,0)$ implies $g_1^{-1} g_2 \in H$.
As the dimension of the coset space is equal to the number of Goldstone boson fields,
the GBs can be identified with elements of $G/H$.
The Goldstone boson fields are said to {\it live} on the coset space
SU(3)$_L \times$ SU(3)$_R$/SU(3)$_V$.

Any $g \in G$ can be decomposed as $g = q h$ with 
$q \in G/H$ and $h \in H$.
The choice of representatives in the coset space $G/H$ is arbitrary.
Possible choices are for example
\beq
g = (g_L, g_R) = (1, g_R g_L^{-1})  (g_L, g_L) \equiv q h
\eeq
or 
\beq
g = (g_L, g_R) = (g_L g_R^{-1},1)  (g_R, g_R) \equiv q' h' \ .
\eeq
If we pick, e.g., the latter choice then the action of $G$ on $G/H$ is given by
\beq
(L,R) (g_L g_R^{-1},1) = (L g_L g_R^{-1}, R) = (L g_L g_R^{-1} R^{-1}, 1) (R,R) \ .
\eeq

The Goldstone bosons are then summarized by the matrix-valued field $U = g_L g_R^{-1}$
which transforms under chiral transformations as
\beq
U(x) \to U'(x) = L U(x) R^{-1} = L U(x) R^{\dagger}  
\eeq
for $L/R \in \mbox{SU(3)}_{L/R}$.
The exponential representation is convenient for $U \in$ SU(3)
\beq
U = \exp \left(\frac{i}{f} \phi^a \lambda^a \right) \ ,
\eeq
where $\lambda^a$ are the generators of SU(3)  
\beq
 \phi = \phi^a \lambda^a = \sqrt{2}
\left(\begin{array}{ccc} \frac{1}{\sqrt{2}} \pi^0 + \frac{1}{\sqrt{6}} \eta &\phantom{+} \pi^+  
& \phantom{+} K^+ \\
\phantom{+}  \pi^- &  -\frac{1}{\sqrt{2}} \pi^0 + \frac{1}{\sqrt{6}} \eta & \phantom{+} K^0\\
\phantom{+} K^- &  \phantom{+} \bar{K}^0 & \phantom{+} - \frac{2}{\sqrt{6}} \eta  \end{array}\right) \ .
\eeq

The chiral effective Lagrangian for QCD is written in terms of the GB fields
which are collected in the matrix-valued field $U$
\beq
\Lagr_{\mbox{\scriptsize{eff}}} = \Lagr_{\mbox{\scriptsize{eff}}} (U, \partial U, \partial^2 U, \ldots) \ .
\eeq
The effective Lagrangian shares the same symmetries with QCD:
$C,P,T$, Lorentz invariance and, in particular, chiral SU(3)$_L \times$ SU(3)$_R$ symmetry.
As outlined above, the chiral Lagrangian is expanded in chiral powers which
are related (in the chiral limit) to the number of derivatives acting on the
GB fields. The chiral power counting of the Lagrangian reads
\beq
\Lagr_{\mbox{\scriptsize{eff}}} = \Lagr_{\mbox{\scriptsize{eff}}}^{(0)} + 
  \Lagr_{\mbox{\scriptsize{eff}}}^{(2)} + \Lagr_{\mbox{\scriptsize{eff}}}^{(4)} + \ldots \ .
\eeq
Only even chiral powers arise since the Lagrangian is a Lorentz scalar
which implies that tensor indices of derivatives appear in pairs.
At each chiral order the effective Lagrangian must be invariant under chiral
SU(3)$_L \times$ SU(3)$_R$ transformations.
At zeroth chiral order this invariance implies that 
$\Lagr_{\mbox{\scriptsize{eff}}}^{(0)}$ can only be a function of 
$U U^\dagger = 1$. This amounts to an irrelevant constant in the Lagrangian
which can be dropped.

At second order, the chiral invariant terms with two derivatives are
\beq
\Lagr_{\mbox{\scriptsize{eff}}}^{(2)} = 
     c_1 \langle \partial_\mu U^\dagger \partial^\mu U \rangle
     +c_2 \langle U^\dagger \Box U \rangle  \ ,
\eeq
where $ \langle \ldots \rangle$ is the trace in flavor space.
The second term can be reduced to the first one by partial integration;
only one term remains at second chiral order
\beq
\Lagr_{\mbox{\scriptsize{eff}}}^{(2)} = 
 c_1 \langle \partial_\mu U^\dagger \partial^\mu U \rangle \ .
\eeq
Since terms of zeroth chiral order have been dropped, the
second chiral order is effectively the leading order (LO).
We note the appearance of a coupling constant $c_1$, a so-called low-energy constant (LEC).
It is fixed by expanding the matrix-valued field $U$ in the GB fields $\phi$
\beq
U = \exp \left(\frac{i}{f} \phi \right) = 1 + \frac{i}{f} \phi - \frac{1}{2 f^2} \phi^2
  + {\cal O}(\phi^3)
\eeq
and requiring the standard kinetic term
\beq
\Lagr_{\mbox{\scriptsize{eff}}}^{(2)} = 
     \frac{1}{2} \partial_\mu \phi^a \partial^\mu \phi^a  + {\cal O}(\phi^4)
\eeq
which yields $c_1 = f^2/4$.

Therefore, the effective Lagrangian at LO reads
\beq
\Lagr_{\mbox{\scriptsize{eff}}}^{(2)} = 
     \frac{f^2}{4} \langle \partial_\mu U^\dagger \partial^\mu U \rangle \ .
\eeq
At leading chiral order there is only one LEC (in the chiral limit) and
chiral symmetry constrains all vertices with increasing number of GB fields
in the LO Lagrangian.

The interpretation of the LEC $f$ can be directly inferred by considering
the Noether axial current of chiral symmetry for $\Lagr_{\mbox{\scriptsize{eff}}}^{(2)} $
\beq
A_\mu^a = 
     i \frac{f^2}{4} \langle \lambda^a \{ \partial_\mu U, U^\dagger \} \rangle \ .
\eeq
Upon comparison with the PCAC hypothesis
\beq
\langle 0 | A_\mu^a (0) | \phi^b (p) \rangle = i  p_\mu \delta^{ab} f_a
\eeq
one confirms that $f$ is the GB decay constant in the chiral limit.

As a first application we are now in a position to predict, e.g., 
$\pi \pi$ scattering at leading chiral order.
The scattering amplitude has the decomposition
\beqa
&& {\cal M} (\pi^a(p_a) \, \pi^b(p_b) \to \pi^c(p_c) \, \pi^d(p_d) ) \no \\ 
 && \quad = \delta^{ab} \delta^{cd} A(s,t,u) 
  +   \delta^{ac} \delta^{bd} A(t,s,u) + \delta^{ad} \delta^{bc} A(u,t,s) \ ,
\eeqa
where $a,b,c,d$ are flavor indices and $s=(p_a + p_b)^2, \, t=(p_a - p_c)^2,\,  u =(p_a - p_d)^2$.
Employing $\Lagr_{\mbox{\scriptsize{eff}}}^{(2)}$ one calculates
\beq  \label{eq:pipi}
A(s,t,u) = \frac{s}{f^2} \ .
\eeq

Up to now, we have worked in chiral limit $m_u, m_d, m_s =0$ where chiral 
symmetry is exact.
In the real world the quark masses do not vanish and introduce an explicit breaking
of chiral symmetry in $\Lagr_{\mbox{\scriptsize{QCD}}}$
\beq
\Lagr_{\mbox{\scriptsize{QCD}}} = \Lagr_{\mbox{\scriptsize{QCD}}}^0 - \bar{q} {\cal M} q =
  \Lagr_{\mbox{\scriptsize{QCD}}}^0 - \bar{q}_R {\cal M} q_L - \bar{q}_L {\cal M} q_R \ ,
\eeq
where $\Lagr_{\mbox{\scriptsize{QCD}}}^0 $ is the massless QCD Lagrangian
and ${\cal M} = \mbox{diag} (m_u,m_d,m_s) $ the light quark mass matrix.
The chiral symmetry breaking 
patterns induced by the light quark masses must be reproduced at the level of the effective field theory.
To this end, we interpret the quark mass matrix as an external scalar source $s$
\beq
\bar{q} {\cal M} q = \bar{q}_L {\cal M} q_R + \bar{q}_R {\cal M}^\dagger q_L  
   \to \bar{q}_L s q_R + \bar{q}_R s^\dagger q_L  \ .
\eeq
The external scalar source $s$ is required to transform under chiral rotations as
\beq
s \to L s R^\dagger \ .
\eeq
Obviously, this leaves the QCD Lagrangian invariant under chiral rotations and implies
that the effective Lagrangian must also remain invariant in the presence of $s$.
Hence, the chiral invariant effective Lagrangian is extended with $s$ as an additional building block
\beq
\Lagr_{\mbox{\scriptsize{eff}}} (U, \partial U, \partial^2 U, \ldots) \to
  \Lagr_{\mbox{\scriptsize{eff}}} (U, \partial U, \partial^2 U, \ldots ; s) \ . 
\eeq
Once the effective Lagrangian is constructed one can go back to the real world
by setting $s = {\cal M}$. In (standard) chiral perturbation 
theory\footnote{We do not consider here the framework of generalized ChPT wherein
${\cal M} = {\cal O}(p)$ \cite{GenChPT}.} 
the chiral counting rule
is $s = {\cal M} = {\cal O}(p^2) $, i.e. the quark masses are booked as second chiral order.
In addition to the derivative expansion the effective Lagrangian is expanded in powers of $s$.
At leading chiral order one obtains
\beqa
\Lagr_{\mbox{\scriptsize{eff}}}^{(2)} & = &
     \frac{f^2}{4} \langle \partial_\mu U^\dagger \partial^\mu U \rangle
     + \frac{f^2}{2} B \langle s U^\dagger + U s^\dagger  \rangle \no \\
     & = &
     \frac{f^2}{4} \langle \partial_\mu U^\dagger \partial^\mu U \rangle
     + \frac{f^2}{2} B \langle {\cal M} (U+ U^\dagger )  \rangle \ .
\eeqa
We have introduced an additional constant $B$ related to explicit chiral symmetry breaking.

The expansion of the symmetry-breaking part in powers of GB fields reads
\beq  \label{eq:massexp}
\frac{f^2}{2} B \langle {\cal M} (U+ U^\dagger ) \rangle = (m_u + m_d + m_s) B f^2 
    - \frac{1}{2} B \langle {\cal M} \phi^2\rangle+ {\cal O}(\phi^4) \ .
\eeq
The first term in this expansion is related to the vacuum expectation values of the scalar quark densities 
\beq
\langle 0 | \bar{q} q | 0 \rangle =  \langle 0 | \frac{\textstyle \partial 
 H_{\mbox{\tiny QCD}}}{\textstyle \partial m_q} | 0 \rangle 
  = - \langle 0 | \frac{\textstyle \partial \Lagr_{\mbox{\tiny eff}}}{\textstyle 
  \partial m_q} | 0 \rangle 
  = - f^2 B + {\cal O} (m_q)
\eeq
and $B$ is the order parameter of spontaneous chiral symmetry breaking.
At lowest order the quark condensates are degenerate
\beq
\langle 0 | \bar{u} u | 0 \rangle = \langle 0 | \bar{d} d | 0 \rangle  =
 \langle 0 | \bar{s} s | 0 \rangle =  - f^2 B  \ .
\eeq

The second term in the expansion in Eq.~(\ref{eq:massexp}) is the mass term for the GB fields
\beq  \label{eq:massrel}
\begin{array}{ll}
m^2_{\pi^+} & = 2 B \hat{m}   \\
  m^2_{\pi^0} & = 2 B \hat{m} + {\cal O} \left[\frac{\textstyle (m_u -m_d)^2}{\textstyle m_s 
                                              - \hat{m}}\right]  \\
  m^2_{K^+} & = B (m_u + m_s)   \\[0.2cm]
  m^2_{K^0} & = B (m_d + m_s)   \\
  m^2_{\eta} & = \frac{2}{3} B ( \hat{m} + 2 m_s) 
       + {\cal O} \left[\frac{\textstyle (m_u -m_d)^2}{\textstyle m_s 
                                              - \hat{m}}\right]  
\end{array}                                              
\eeq
with $\hat{m}= \frac{1}{2}(m_u + m_d)$.
The terms proportional to $(m_u -m_d)^2$ are due to $\pi^0$-$\eta$ mixing.
They constitute tiny corrections and are usually neglected.
The mass relations in Eq.~(\ref{eq:massrel}) explain why one counts $s= {\cal O}(p^2) $ (taking $B$
to be of zeroth chiral order).
For finite light quark masses we see that the GBs are not massless any longer 
(sometimes they are referred to as {\it pseudo}-GBs).

From the previous equalities one obtains immediately the Gell-Mann--Oakes--Renner relations
\beq
\begin{array}{ll}
  f_\pi^2 m^2_{\pi} & = - 2 \hat{m}  \langle 0 | \bar{q} q | 0 \rangle + {\cal O}(m_q^2) \\[0.2cm]
  f_K^2 m^2_{K} & = - ( \hat{m} + m_s)  \langle 0 | \bar{q} q | 0 \rangle + {\cal O}(m_q^2) \\[0.2cm]
  f_\eta^2 m^2_{\eta} & = -\frac{2}{3} 
                     ( \hat{m} + 2 m_s) \langle 0 | \bar{q} q | 0 \rangle + {\cal O}(m_q^2)\\[0.2cm]
  \end{array}
\eeq
and the Gell-Mann--Okubo mass relation
\beq
3 m^2_{\eta} =   4m^2_{K} - m^2_{\pi} \ .
\eeq

To leading chiral order, the strong interactions at low energies are characterized by the two scales
$f$ and $B$ with
\beq
\begin{array}{rl}
  f \simeq f_\pi \ & \simeq 93 \MeV \\[0.2cm]
  B \ & \simeq 1800 \MeV  \ ,
\end{array}
\eeq
where the value of $B$ has been extracted from the sum-rule value 
$\langle 0 | \bar{q} q | 0 \rangle = -(250 \MeV)^3$ \cite{SVZ}.

One important goal of chiral perturbation theory is to
extract the light quark mass ratios from phenomenology.
Note that the absolute values of the quark masses cannot be extracted since they depend
on the QCD renormalization scale, but the QCD scale dependence cancels out in quark mass ratios.
For example, from the leading order mass relations one obtains
\beq
\frac{m_u}{m_d} = \frac{(m^2_{K^+}- m^2_{K^0})_{\mbox{\tiny QCD}} + m^2_{\pi^0}}{
           (m^2_{K^0}- m^2_{K^+})_{\mbox{\tiny QCD}}+ m^2_{\pi^0}} \ ,
\eeq
where the tiny contribution from $\pi^0$-$\eta$ mixing 
and electromagnetic effects in the kaon mass difference have been neglected.
Next, we employ Dashen's theorem \cite{Dash} which asserts that
electromagnetic contributions to the  mass splittings $m^2_{K^+}- m^2_{K^0}$
and $m^2_{\pi^+}- m^2_{\pi^0}$ are identical in the chiral limit
\beq
\begin{array}{ll}
  (m^2_{K^+}- m^2_{K^0})_{\mbox{\scriptsize QCD}} & = (m^2_{K^+}- m^2_{K^0})_{\mbox{\scriptsize phys}} 
                      - (m^2_{K^+}- m^2_{K^0})_{\mbox{\scriptsize  em}} \\[0.2cm]
   & = (m^2_{K^+}- m^2_{K^0})_{\mbox{\scriptsize phys}} 
                      - (m^2_{\pi^+}- m^2_{\pi^0})_{\mbox{\scriptsize  em}} \\[0.2cm]
    & = (m^2_{K^+}- m^2_{K^0})_{\mbox{\scriptsize phys}} 
                      - (m^2_{\pi^+}- m^2_{\pi^0})_{\mbox{\scriptsize  phys}} \ .
   \end{array}
\eeq
The last equation is valid
since the pion mass difference is (almost) entirely due to electromagnetic contributions.
Insertion of this identity in the quark mass ratio yields \cite{Wei1}
\beq
\frac{m_u}{m_d} = \frac{m^2_{K^+}- m^2_{K^0} + 2 m^2_{\pi^0}-m^2_{\pi^+}}{
           m^2_{K^0}- m^2_{K^+}+ m^2_{\pi^+}} 
        = 0.55 \ .
\eeq
At leading order, the possibility $m_u =0$ which would lead to a solution 
of the strong $CP$ problem is excluded.
In a similar manner one finds \cite{Wei1}
\beq
\frac{m_s}{m_d} = \frac{m^2_{K^0}+ m^2_{K^+} - m^2_{\pi^0}}{
           m^2_{K^0}- m^2_{K^+}+ m^2_{\pi^+}}  
        = 20.1 \ .
\eeq
These leading order ratios are, of course, subject to higher order corrections.

We note in passing that due to the inclusion of 
finite quark masses the $\pi \pi$ scattering amplitude, Eq.~(\ref{eq:pipi}), is modified
according to
\beq
A(s,t,u) = \frac{s - m_\pi^2}{f^2} = \frac{s - m_\pi^2}{f^2_\pi} + {\cal O}(p^4) \ .
\eeq

In the remainder of this section we extend the effective field theoretical
formalism to include external vector and axial-vector fields.
This is necessary for the calculation of Green functions of quark currents,
e.g., for the electromagnetic form factor of the pion
\beq
\langle \pi(p') | J_\mu^{\mbox{\scriptsize{em}}}  | \pi(p) \rangle =
  \langle \pi(p') | \bar{q} \gamma_\mu Q q  | \pi(p) \rangle \ ,
\eeq
where $Q = \frac{1}{3} \mbox{diag} (2,-1,-1)$ is the quark charge matrix.
Green functions of quark currents are conveniently calculated by applying 
the external field method. (As a matter of fact, we have 
already encountered this technique for the scalar quark densities $\langle 0 |\bar{q} q | 0\rangle $
which are obtained by differentiating with respect to the external field $s$.)

To this aim, one extends $\Lagr_{\mbox{\scriptsize{QCD}}}$ in the presence of external fields
\beq  \label{eq:LagrQCDext}
\Lagr_{\mbox{\scriptsize{QCD}}} = \Lagr_{\mbox{\scriptsize{QCD}}}^{0}
    + \bar{q} \gamma_\mu (v^\mu + \gamma_5 a^\mu) q
    - \bar{q} (s + i \gamma_5 p) q \ ,
\eeq
where $v^\mu$, $a^\mu$, $s$, and $p$ are hermitian matrices denoting the external 
vector, axial-vector, scalar, and pseudoscalar fields, respectively.
The coupling of external photon fields ${\cal A}_\mu$ to the quarks, e.g., is given by the
prescription $v_\mu = -e Q {\cal A}_\mu$.
Green functions of electromagnetic currents are then obtained by functional derivatives
of the generating functional $Z[v,a,s,p]$ with respect to  
${\cal A}_\mu$.
Couplings to weak currents are also collected in $v_\mu$ and $a_\mu$.

Due to the inclusion of electromagnetic and weak currents 
global chiral symmetry is promoted to a local one.  
The Lagrangian $\Lagr_{\mbox{\scriptsize{QCD}}}$ in Eq.~(\ref{eq:LagrQCDext})
remains invariant under the local chiral 
SU(3)$_L \times$SU(3)$_R$ transformations 
\beq
\begin{array}{rl}
  q_R \ & \to \ R q_R \\[0.2cm]
  q_L \ & \to \ L q_L \\[0.2cm]
  r_\mu = v_\mu + a_\mu \ & \to \ R r_\mu R^\dagger + i R \partial_\mu R^\dagger \\[0.2cm]
  l_\mu = v_\mu - a_\mu \ & \to \ L l_\mu L^\dagger + i L \partial_\mu L^\dagger \\[0.2cm]
  s + ip \ & \to \ L (s + ip ) R^\dagger \ .
  \end{array}
\eeq
Local chiral symmetry must also be maintained by the extended effective Lagrangian
in the presence of external fields
\beq
\Lagr_{\mbox{\scriptsize{eff}}} (U, \partial U, \partial^2 U, \ldots) \to
  \Lagr_{\mbox{\scriptsize{eff}}} (U, \partial U, \partial^2 U, \ldots ; v,a,s,p)  \ .
\eeq

The local nature of chiral symmetry requires the introduction of a covariant derivative
\beq
\partial_\mu U \to D_\mu U = \partial_\mu U - i l_\mu U + i U r_\mu
\eeq
which transforms under chiral rotations as
\beq
D_\mu U \to  L (D_\mu U ) R^\dagger \ .
\eeq
In addition, the vector and axial-vector fields $v_\mu$, $a_\mu$ enter the effective Lagrangian 
also through the non-Abelian field strength tensors, 
$R_{\mu \nu}, L_{\mu \nu}$,
\beq
\begin{array}{l}
R_{\mu \nu} = \partial_\mu r_\nu - \partial_\nu r_\mu - i [r_\mu,r_\nu]\\[0.2cm]
  L_{\mu \nu} = \partial_\mu l_\nu - \partial_\nu l_\mu - i [l_\mu,l_\nu]
\end{array} 
\eeq
with the transformation properties
\beq
R_{\mu \nu} \to R \, R_{\mu \nu} \, R^\dagger \ , \qquad
  L_{\mu \nu} \to L \, L_{\mu \nu} \, L^\dagger \ .
\eeq
%

%%%%%%%%%%%%%%%%%%%%%%%%%%%%%%%%%%%%%%%%%%%%%%%%%%%%%%%%%%%%%%%%%%%%%%%%%%%
\section{Higher orders and loops}
\label{sec:HiOrdLoop}

In the previous section we have outlined the construction principles
for the chiral effective Lagrangian. The effective Lagrangian is ordered
in powers of Goldstone boson momenta and masses, a chiral power counting scheme
is established. The leading order effective Lagrangian 
$\Lagr_{\mbox{\scriptsize{eff}}}^{(2)}$ was constructed
explicitly and sample calculations have been performed at the leading tree level order.
In this section we focus on the inclusion of next-to-leading order (NLO) contributions.
On the one hand, these are the tree level diagrams with vertices from 
$\Lagr_{\mbox{\scriptsize{eff}}}^{(4)}$. On the other hand, loop diagrams
with vertices from $\Lagr_{\mbox{\scriptsize{eff}}}^{(2)}$ must be taken into account.

The chiral expansion of the effective Lagrangian reads
\beq
\Lagr_{\mbox{\scriptsize{eff}}} = 
  \Lagr_{\mbox{\scriptsize{eff}}}^{(2)} + \Lagr_{\mbox{\scriptsize{eff}}}^{(4)} + \ldots \ .
\eeq
The most general effective Lagrangian at fourth chiral order,
$\Lagr_{\mbox{\scriptsize{eff}}}^{(4)}$, which respects the relevant symmetries of QCD 
is \cite{GL2}\footnote{We have omitted two terms that are not accessible from experiment and only serve
the purpose of absorbing loop divergences.}
\beq
\begin{array}{ll}
  \Lagr_{\mbox{\scriptsize{eff}}}^{(4)} & = L_1 \langle D_\mu U^\dagger D^\mu U \rangle ^2
   + L_2 \langle D_\mu U^\dagger D_\nu U \rangle \langle D^\mu U^\dagger D^\nu U \rangle \\[0.2cm]
  & + \, L_3 \langle D_\mu U^\dagger D^\mu U D_\nu U^\dagger D^\nu U \rangle 
    + L_4 \langle D_\mu U^\dagger D^\mu U \rangle \langle \chi U^\dagger 
                                                  + U \chi^\dagger \rangle\\[0.2cm]
  & + L_5 \langle D_\mu U^\dagger D^\mu U ( \chi^\dagger U + U^\dagger  \chi) \rangle
    + L_6 \langle \chi U^\dagger + U \chi^\dagger \rangle ^2\\[0.2cm]
  & + L_7 \langle \chi U^\dagger - U \chi^\dagger \rangle ^2
    + L_8 \langle \chi U^\dagger \chi U^\dagger + U \chi^\dagger U \chi^\dagger \rangle \\[0.2cm]
  & - i L_9 \langle R^{\mu \nu} D_\mu U^\dagger D_\nu U 
                 +   L^{\mu \nu} D_\mu U D_\nu U^\dagger  \rangle 
    + L_{10} \langle R_{\mu \nu}  U^\dagger L^{\mu \nu}  U  \rangle 
  \end{array}
\eeq
with $\chi \equiv 2 B (s +ip)$.
Ten low-energy constants enter with unknown values which must be determined
from phenomenology.
At next-to-next-to-leading order even 90 unknown LECs enter \cite{BCE2}.
Due to the proliferation
of unknown couplings in the effective Lagrangian ChPT seems to lose its predictive power.
However, the situation is ameliorated since only a few LECs contribute to a particular process.
There is no need to determine all LECs of the effective Lagrangian for a given process.
Moreover, the contributions from higher chiral orders are suppressed as they appear
with additional powers of small momenta or Goldstone boson masses.
Therefore in practice only the LECs from lower orders are required.

\begin{figure}[ht]
\centering
        \includegraphics[width=2.cm]{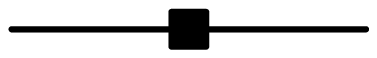}
\caption{Tree diagram contribution to the self-energy of the Goldstone bosons.
         The square denotes a vertex from $\Lagr_{\mbox{\scriptsize{eff}}}^{(4)}$.}
\label{fig:L4mass}
\end{figure}

\begin{figure}[b]
\centering
        \includegraphics[width=1.5cm]{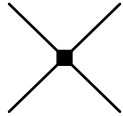}
\caption{Tree diagram contribution to $\pi \pi$ scattering at order order ${\cal O}(p^4)$.
         The square denotes a vertex from $\Lagr_{\mbox{\scriptsize{eff}}}^{(4)}$.}
\label{fig:L4scatt}
\end{figure}

At fourth chiral order, contributions from $\Lagr_{\mbox{\scriptsize{eff}}}^{(4)}$
to the pseudoscalar meson masses stem solely from tree diagrams, see Fig.~\ref{fig:L4mass}.
Corrections from 
$\Lagr_{\mbox{\scriptsize{eff}}}^{(4)}$ to $\pi \pi$ scattering
at ${\cal O}(p^4)$ also contribute through tree diagrams, Fig.~\ref{fig:L4scatt}.

As already indicated above, tree diagrams are not the whole story.
At next-to-leading order loop diagrams must be included as well.
The loops also restore unitarity of the $S$-matrix 
which can be seen as follows.
Unitarity of the $S$-matrix implies
\beq
S S^\dagger =1 \ .
\eeq
Inserting the usual decomposition $S= 1 + iT$ yields
\beq
-i (T- T^\dagger) = T^\dagger T \ .
\eeq
Since $T$ is symmetric due to time reversal invariance the last relation simplifies to
\beq   \label{eq:unit}
2 \, \mbox{Im} \, T = T^\dagger T \ . 
\eeq
Tree diagrams do not have an imaginary piece and thus cannot satisfy Eq.~(\ref{eq:unit}), i.e.,
unitarity is violated if only tree diagrams are taken into account.
However, unitarity of the $S$-matrix is perturbatively restored by including loop diagrams
which have imaginary parts.

\begin{figure}[ht]
\centering
        \includegraphics[width=2.6cm]{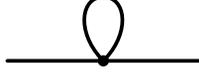}
\caption{Tadpole contribution to the self-energy of the Goldstone bosons.
         The vertex stems from $\Lagr_{\mbox{\scriptsize{eff}}}^{(2)}$.}
\label{fig:tadpole}
\end{figure}

As a simple example of a loop calculation, we sketch the evaluation of the one-loop contributions
to the GB self-energies.
To this end, we expand the $U$ fields in $\Lagr_{\mbox{\scriptsize{eff}}}^{(2)}$
\beq
\begin{array}{ll}
   \Lagr_{\mbox{\scriptsize{eff}}}^{(2)} & = 
     \frac{f^2}{4} \langle D_\mu U^\dagger D^\mu U \rangle
     + \frac{f^2}{2} B \langle {\cal M} ( U^\dagger + U )  \rangle \\[0.2cm]
     & = \Lagr_{\mbox{\scriptsize{kin}}} + \Lagr_{\mbox{\scriptsize{int}}} \\[0.2cm]
     & = 
      \frac{1}{4} \langle \partial_\mu \phi \partial^\mu \phi \rangle
     - \frac{1}{2} B \langle {\cal M} \phi^2  \rangle \\[0.2cm]
     & \quad +  \frac{1}{48 f^2} \langle [\partial_\mu \phi, \phi] [\partial^\mu \phi,\phi] \rangle
     + \frac{1}{24 f^2} B \langle {\cal M} \phi^4  \rangle + \ldots \quad .
    \end{array}
\eeq
Interaction vertices with four GB fields arise which
yield contributions to the meson masses via tadpoles, see  Fig.~\ref{fig:tadpole}.

The calculation of the tadpole diagram involves the loop integration 
\beq
\int \frac{d^4 k}{(2 \pi)^4} \frac{i}{k^2 - m_\phi^2+ i \epsilon } \ ,
\eeq
where $ m_\phi $ is the GB mass in the loop.
This loop integral is quadratically ultraviolet divergent and a
regularization scheme is necessary which maintains the symmetries of the theory, in particular
chiral symmetry.
One must be careful when employing cutoff regularization schemes since
these introduce an additional massive cutoff scale which may spoil chiral symmetry \cite{GJLW}.
In this respect, a convenient mass-independent regularization scheme is provided by 
dimensional regularization.
The basic idea of dimensional regularization is as follows.
The dimension of the loop integration is changed to an arbitrary dimension 
$d \in \mathds{R}$.
For $d$ small enough the integral remains ultraviolet finite and can be performed
explicitly.
Afterwards, the analytic continuation back to $d=4$ dimensions is performed.

To be more explicit, consider the tadpole integral in $d$ dimensions
\beq
I = \mu^{4-d} \int \frac{d^d k}{(2 \pi)^d} \frac{i}{k^2 - m_\phi^2 + i \epsilon} \ .
\eeq
The regularization scale $\mu$ has been introduced to maintain the mass dimension of $I$
for arbitrary space-time dimension $d$.
The calculation of $I$ is performed in a straightforward manner
\beq
\begin{array}{ll}
  I & =  \mu^{4-d} {\displaystyle\int \frac{d^d k}{(2 \pi)^d} \frac{i}{k^2 - m_\phi^2
                                          + i \epsilon}} 
   = {\displaystyle \mu^{4-d} \int \frac{d^d k_{\mbox{\tiny E}}}{(2 \pi)^d} 
    \frac{1}{k^2_{\mbox{\tiny E}} + m_\phi^2}} \\[0.4cm]   
  & = {\displaystyle \mu^{4-d} \int \frac{d^d k_{\mbox{\tiny E}}}{(2 \pi)^d} 
      \int_0^\infty d \lambda \ \exp \left\{- \lambda \left( k^2_{\mbox{\tiny E}} + m_\phi^2 \right)
      \right\}} \\[0.4cm]    
  & = {\displaystyle \frac{\mu^{4-d} }{(2 \pi)^d} \int_0^\infty d \lambda 
                                 \ \exp \left(- \lambda  m_\phi^2 \right)
        \int d^d k_{\mbox{\tiny E}} \ \exp \left(- \lambda k^2_{\mbox{\tiny E}} \right) }\\[0.4cm]   
  & = {\displaystyle \frac{\mu^{4-d} }{(2 \pi)^d} \int_0^\infty d \lambda 
                                 \ \exp \left(- \lambda  m_\phi^2 \right) 
                                 \left(\frac{\pi}{\lambda}\right)^{d/2}  }\\[0.4cm]   
  & = {\displaystyle (4 \pi)^{- \frac{d}{2}} m_\phi^2 \left(\frac{m_\phi}{\mu}\right)^{d-4} 
                   \Gamma(1- \frac{d}{2} )   }  \ .
  \end{array}
\eeq
A Wick rotation to Euclidean space has been performed in the first line.
Expanding the result in $d-4$ yields
\beq
I  = m_\phi^2 \left\{ 2 L + \frac{1}{16 \pi^2} \ln \left( \frac{m_\phi^2}{\mu^2} \right) \right\}
  + {\cal O}(d-4)
\eeq
with the divergent piece
\beq
L = \frac{\mu^{d-4}}{16 \pi^2} \left[ \frac{1}{d-4} - \frac{1}{2} 
                            \left( \ln 4 \pi +1 + \Gamma' (1)   \right) \right] \ .
\eeq
The tadpole integral $I$ is decomposed into two parts: a 
non-analytic term from the GB infrared singularities proportional to the ``chiral log''
$\ln (m_\phi^2/\mu^2)$ and a piece $L$ which diverges for $ d \to 4$.

A complete one-loop calculation for GB self-energies with vertices from 
$\Lagr_{\mbox{\scriptsize{eff}}}^{(2)}$ reveals that divergent parts proportional to $L$
are of fourth chiral order and analytic in the quark masses and GB momenta.
Hence, the divergent components can be compensated by renormalizing the LECs of fourth chiral order
\beq
L_i = L_i^r + \lambda_i L
\eeq
such that the sum of loops and LECs (including fourth chiral order) remains finite
for $d\to 4$ and can be evaluated in four space-time dimensions.

The above considerations for one-loop contributions to the self-energy
can be generalized to other processes and higher loops. 
Consider a connected $L$-loop diagram for a given physical process.
The chiral order (or chiral dimension) $D$ of a loop integral counts dimensions of GB masses and momenta
(utilizing a mass-independent regularization scheme such as, e.g., dimensional regularization)
and is given by 
\beq  \label{eq:chidim}
D = 2 + 2 L +  \sum_k (k-2) N_k \ ,
\eeq
where $L$ is the number of loops and $N_k$ the number of vertices of order ${\cal O}(p^k)$.

Vertices from the lowest order Lagrangian $\Lagr_{\mbox{\scriptsize{eff}}}^{(2)}$
do not alter the chiral dimension, while vertices from higher orders increase $D$.
Each loop also increases the chiral dimension, i.e. the loopwise expansion
corresponds with the chiral expansion.
From Eq.~(\ref{eq:chidim}) one infers that at leading chiral order only
tree diagrams with vertices of $\Lagr_{\mbox{\scriptsize{eff}}}^{(2)}$ contribute.
At next-to-leading order both tree diagrams with exactly one fourth order vertex
and one-loop diagrams with vertices of second chiral order contribute.
Two-loop diagrams start contributing at sixth chiral order and so forth.

\begin{figure}[ht]
\centering
        \includegraphics[width=1.5cm]{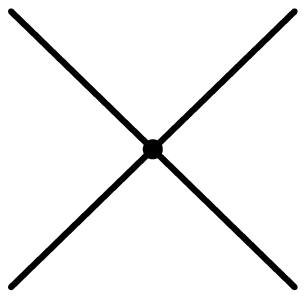}
\caption{Leading ${\cal O}(p^2)$ contribution to $\pi \pi$ scattering.
         The vertex stems from $\Lagr_{\mbox{\scriptsize{eff}}}^{(2)}$.}
\label{fig:L2scatt}
\end{figure}

\begin{figure}[b]
\centering
        \parbox{2.4cm}{\centering \includegraphics[width=2.3cm]{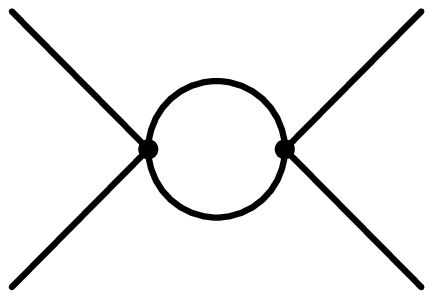}} \qquad
  \parbox{2.4cm}{\centering \includegraphics[width=1.9cm]{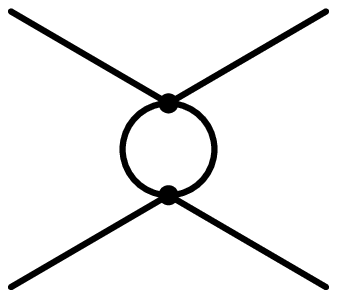}} \qquad
  \parbox{2.4cm}{\centering \includegraphics[width=1.9cm]{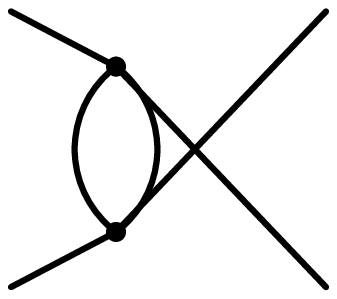}}
\caption{One-loop diagrams for $\pi \pi$ scattering with vertices from 
                  $\Lagr_{\mbox{\scriptsize{eff}}}^{(2)}$.}
\label{fig:scattloop}
\end{figure}

As an explicit example of the chiral expansion we consider the Feynman diagrams
for $\pi \pi$ scattering up to one-loop order.
At lowest order, ${\cal O}(p^2)$, a tree diagram with a lowest order vertex
contributes, see Fig.~\ref{fig:L2scatt}.
At next-to-leading order, ${\cal O}(p^4)$,
both tree diagrams with vertices from $\Lagr_{\mbox{\scriptsize{eff}}}^{(4)}$ 
(Fig.~\ref{fig:L4scatt}) and one-loop diagrams 
with lowest order vertices (Fig.~\ref{fig:scattloop}) contribute.
The corresponding loop integrals are of the type
\beq
\int \frac{d^d l}{(2 \pi)^d} \frac{p_1 \cdot p_2 \ \ p_3 \cdot p_4}{[l^2 - m_a^2 + i \epsilon] 
          \ [(l+ p_1 + p_2)^2 - m_b^2 + i \epsilon]}  \ .
\eeq

The construction of the most general chiral effective Lagrangian compliant with the symmetries
and usage of a regularization scheme which respects these symmetries
guarantees that appearing divergences can be absorbed in the LECs.
Therefore, quantum effects do not violate chiral symmetry.
For a mass-independent regularization scheme the occuring divergences will be of 
higher chiral order than the involved vertices.
In this way, a systematic renormalization procedure emerges.
The complete renormalization program at one- and two-loop order has been carried out
by employing a background field method and heat kernel techniques \cite{GL1, GL2,BCE}.

The chiral expansion is ordered in increasing powers of GB masses
and momenta.
Each additional loop produces a factor of $1/f^2$ from the expansion of $U = \exp (i \phi /f)$ 
multiplied by the loop factor $1/(4 \pi)^2$.
Since each loop increases the chiral dimensionality $D$ by two,
inverse powers of the scale $4 \pi f \simeq 4 \pi f_\pi \simeq 1.2 \GeV$
enter in the chiral expansion.
The scale of spontaneous chiral symmetry breakdown is thus 
\beq
\Lambda_\chi = 4 \pi f_\pi \simeq 1.2 \GeV \ .
\eeq
Moreover, variation of the regularization scale $\mu$ allows for shuffling finite contributions
from loops to LECs and vice versa. This suggests that contributions from 
renormalized LECs have roughly the same order of magnitude and, thus, the
same expansion scale.
Based on this naive estimate, the natural expansion parameter of the chiral amplitudes
is expected to be
\beq
\frac{m_\eta^2}{16 \pi^2 f_\pi^2} \simeq 0.22  \ .
\eeq
However, in practice substantial higher order corrections which are not in accord with this rule
may occur in the chiral expansion.

Let us focus now on the values of the renormalized constants which are not 
fixed by chiral symmetry. They must be calculated either directly from QCD
or fixed from experimental data. In the first case one tries to extract
the values of the LECs from lattice QCD simulations, but 
only a smaller set of LECs has been extracted so far using this procedure,
see e.g. \cite{LEClat}.
The second option, i.e. the direct comparison of the chiral amplitudes
with available experimental data, has been applied more extensively.
The phenomenological values and sources for the renormalized coupling constants $L_i^r$
at the regularization scale $\mu = M_\rho$ are presented in Table~\ref{tab:LECs}.
\begin{table}
\centering
\begin{tabular}{|c|r|l|}
  \hline
  \ $i$ \ & \ $L_i^r(\mu= M_\rho) \times 10^3$ \ & \ source\\[0.15cm]
  \hline
  \ 1 \ & $0.4 \pm 0.3$ \ & \  $\pi \pi \to \pi \pi , K_{e4}$ \ \\[0.1cm]
  \ 2 \ & $1.35 \pm 0.3$ \ & \  $\pi \pi \to \pi \pi , K_{e4}$ \ \\[0.1cm]
  \ 3 \ & $-3.5 \pm 1.1$ \ & \  $\pi \pi \to \pi \pi , K_{e4}$ \ \\[0.1cm]
  \ 4 \ & $-0.3 \pm 0.5$ \ & \  Zweig rule \ \\[0.1cm]
  \ 5 \ & $1.4 \pm 0.5$ \ & \  $f_K/f_\pi$ \ \\[0.1cm]
  \ 6 \ & $-0.2 \pm 0.3$ \ & \  Zweig rule \ \\[0.1cm]
  \ 7 \ & $-0.4 \pm 0.2$ \ & \  Gell-Mann--Okubo \ \\[0.1cm]
  \ 8 \ & $0.9 \pm 0.3$ \ & \  $m_{K^0}^2 -  m_{K^+}^2$ \ \\[0.1cm]
  \ 9 \ & $6.9 \pm 0.7$ \ & \  $\langle r^2 \rangle_V^\pi$ \ \\[0.1cm]
  \ 10 \ & $-5.5 \pm 0.7$ \ & \  $\pi \to e \nu \gamma$ \\
  \hline
\end{tabular}
\caption{Phenomenological values and sources for the renormalized coupling 
         constants $L_i^r(\mu= M_\rho)$ \cite{BEG}.}
\label{tab:LECs}
\end{table}
Once the LECs are determined one can predict further observables at the same chiral order.

Consider, e.g., the extraction of $L_5$.
The LEC $L_5$ governs SU(3) breaking in the GB decay constants
\beq
\frac{f_K}{f_\pi} = 1 + \frac{4 L_5^r}{f^2} (m_K^2 - m_\pi^2) + \mbox{chiral logs} \ .
\eeq
Comparison with the experimental value \cite{pdg} 
\beq
\left(\frac{f_K}{f_\pi}\right)_{\mbox{\scriptsize exp}} = 1.22
\eeq
fixes $L_5^r$.

\begin{figure}[ht]
\centering
\parbox{3.cm}{\centering \includegraphics[width=1.8cm]{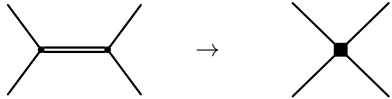}} $\rightarrow$
  \parbox{3.cm}{\centering \includegraphics[width=1.3cm]{L4scatt.eps}}
\caption{In the static limit of heavy resonance masses (represented by the double line)
         vertices of the chiral effective Lagrangian are produced.}
\label{fig:resex}
\end{figure}

The values of the $L_i$ can be understood in terms of meson resonance exchange \cite{reso}.
For resonance masses considerably larger than the involved momenta, $M_R^2 \gg p^2$,
the resonance propagators shrink to a single point
and produce contact interactions of the chiral Lagrangian
\beq
\frac{1}{p^2 - M_R^2} \to - \frac{1}{M_R^2} + {\cal O}(p^2) \ .
\eeq
Diagrammatically this is depicted in Fig.~\ref{fig:resex}.
Resonance saturation is based on the assumption that 
the $L_i^r(\mu= M_\rho)$ are practically saturated by resonance exchange (chiral duality).
This principle appears to work very well, see Table~\ref{tab:LECsres}.
\begin{table}
\centering
\begin{tabular}{|c|r|l|}
  \hline
  \ $i$ \ & \ $L_i^r(\mu= M_\rho) \times 10^3$ \ & \ $L_i^{\mbox{\scriptsize res}}  \times 10^3$ \\[0.15cm]
  \hline
  \ 1 \ & $0.4 \pm 0.3$ \ & \  $0.6$ \ \\[0.1cm]
  \ 2 \ & $1.35 \pm 0.3$ \ & \ $1.2$  \ \\[0.1cm]
  \ 3 \ & $-3.5 \pm 1.1$ \ & \  $-3.0$ \ \\[0.1cm]
  \ 4 \ & $-0.3 \pm 0.5$ \ & \  $0.0$ \ \\[0.1cm]
  \ 5 \ & $1.4 \pm 0.5$ \ & \  $1.4$ \ \\[0.1cm]
  \ 6 \ & $-0.2 \pm 0.3$ \ & \ $0.0$  \ \\[0.1cm]
  \ 7 \ & $-0.4 \pm 0.2$ \ & \  $-0.3$ \ \\[0.1cm]
  \ 8 \ & $0.9 \pm 0.3$ \ & \  $0.9$ \ \\[0.1cm]
  \ 9 \ & $6.9 \pm 0.7$ \ & \ $6.9$  \ \\[0.1cm]
  \ 10 \ & $-5.5 \pm 0.7$ \ & \ $-6.0$  \\
  \hline
\end{tabular}
\caption{Phenomenological values $L_i^r(\mu= M_\rho)$ in comparison with the
        values $L_i^{\mbox{\scriptsize res}} $ from resonance saturation \cite{reso}.}
\label{tab:LECsres}
\end{table}
One also observes that whenever spin-1 resonances contribute they dominate the resonance exchange 
(chiral vector meson dominance).

After setting up the NLO Lagrangian we are now in a position to calculate
the light quark mass ratios at next-to-leading order.
The ${\cal O}(p^4)$ expressions for the light quark mass ratios are
\beq
\frac{m_K^2}{m_\pi^2} = \frac{m_s + \hat{m}}{m_u + m_d} \left[ 1 + \Delta_M + {\cal O}(m_q^2) \right] 
\eeq
and
\beq
\frac{(m_{K^0}^2-m_{K^+}^2)_{\mbox{\tiny QCD}}}{m_K^2 -m_\pi^2} = 
    \frac{m_d - m_u }{m_s - \hat{m}} \left[ 1 + \Delta_M + {\cal O}(m_q^2) \right]
\eeq
with the same chiral correction
\beq
\Delta_M =  \frac{8}{f^2} [ 2 L_8^r - L_5^r ] \ (m_K^2 - m_\pi^2) + \mbox{chiral logs} \ .
\eeq
Combining both results one obtains the parameter-free relation
\beq
Q^2 \equiv \frac{m_s^2 - \hat{m}^2}{m_d^2 - m_u^2 } =
  \frac{m_K^2}{m_\pi^2}  \frac{m^2_K- m^2_\pi}{m^2_{K^0}-m^2_{K^+}  + m^2_{\pi^+} - m^2_{\pi^0}} \ ,
\eeq
where Dashen's theorem \cite{Dash} has been employed
\beq
(m_{K^0}^2-m_{K^+}^2)_{\mbox{\tiny QCD}} =  m^2_{K^0}-m^2_{K^+}  + m^2_{\pi^+} - m^2_{\pi^0} \ .
\eeq
This yields
\beq
Q_{\mbox{\tiny Dashen}} \simeq 24.1 \ .
\eeq
It is argued in the literature that chiral corrections to Dashen's theorem may
decrease $Q$ by up to 10\% \cite{Dashcorr}.
In fact, dispersive calculations of the decay $\eta \to 3 \pi$ also suggest a slightly lower value
\cite{Qval}, but the exact value for $Q$ is still under lively discussion.\\

\noindent
We close this section with two remarks.\\

\paragraph{SU(2) chiral perturbation theory}
In this presentation, we have assumed that the strange quark mass $m_s$
is light and in the chiral regime. The appropriate framework then is SU(3) ChPT.
Since the group SU(2) is a subgroup of SU(3), the SU(3) Lagrangian is
also valid for chiral SU(2) relevant for the pions $(\pi^\pm ,\pi^0)$.
If, on the other hand, $m_s$ is considered to be large, 
kaons and the eta are heavy particles and may be integrated out
such that only pionic degrees of freedom remain \cite{GL2}. The result is a
pure SU(2) chiral Lagrangian with less LECs \cite{GL1}.\\

\paragraph{Axial U(1) anomaly}
Finally, we would like to comment on the axial U(1) anomaly of the strong interactions.
At the classical level, the QCD Lagrangian exhibits a U(3)$_L \times$U(3)$_R$ symmetry
\beq
\mbox{U(3)}_L \times \mbox{U(3)}_R = \mbox{SU(3)}_L \times \mbox{SU(3)}_R \times 
  \mbox{U(1)}_{V} \times \mbox{U(1)}_{A}
\eeq
which comprises the SU(3)$_L \times$SU(3)$_R$ symmetry considered so far.
Under the transformations
\beq
q_R \rightarrow R \, q_R, \quad q_L \rightarrow L \, 
  q_L, \qquad R,L \in \mbox{U(3)}
\eeq
the classical QCD Lagrangian remains invariant.
The U(1)$_V$ symmetry corresponds with baryon number conservation and is usually
neglected.
In contrast, the U(1)$_A$ symmetry of the classical Lagrangian is broken at 
the quantum level. 
Under axial U(1) transformations the path integral picks up an additional contribution
from the fermion determinant such that the full quantum theory does not
exhibit the axial U(1) symmetry \cite{anomalies}. This is the so-called axial U(1) anomaly.

If the axial U(1) symmetry were not broken by the anomaly, it would imply
the  existence of another isoscalar $0^-$ Goldstone boson with a small mass \cite{Wein}
\beq
m \le \sqrt{3} m_\pi \ .
\eeq
The lightest pseudoscalar meson and singlet under SU(3)$_V$ is the  $\eta'$
with a mass
\beq
m_{\eta'}  = 958 \MeV \sim \Lambda_\chi \gg m_{\pi, K, \eta}
\eeq
close to the scale of spontaneous chiral symmetry breakdown, $\Lambda_\chi$.
The $\eta'$ is not a Goldstone boson and remains massive in the chiral limit.
Hence, it is not included explicitly 
in conventional ChPT although its effects are hidden in coupling constants 
of the Lagrangian \cite{GL2}.

However, in the limit $N_c \rightarrow \infty$, where $N_c$ is the number of colors,
the axial U(1) anomaly vanishes and a nonet of Goldstone bosons is generated,
$(\pi, K,\eta, \eta')$, which now includes the $\eta'$ as ninth Goldstone boson
with a mass comparable to the other GBs. Stated differently, the
extra mass of the $\eta'$ is induced by the axial anomaly. 
Starting from the large $N_c$ limit it is possible to include the $\eta'$ 
systematically in the effective theory \cite{GL2, etap}. The chiral amplitudes can be expanded
simultaneously both in chiral powers and powers of $1/N_c$ which establishes 
a rigorous counting scheme \cite{largeNc}.
Besides, by adding the $\eta'$ in the effective Lagrangian
strong $CP$ violation is automatically included and allows for the calculation of 
strong $CP$ violating effects, such as the electric dipole moment of the neutron
\cite{EDM}.

\section*{Acknowledgements}
It is a pleasure to thank the organizers of the summer school for creating a very pleasant 
and stimulating atmosphere in a beautiful environment.
I also thank Peter Bruns and Robin Ni{\ss}ler for a careful reading of the manuscript.
%

%
%
% Non-BibTeX users please follow the syntax
% the syntax of "referenc.tex" for your own citations
%%%%%%%%%%%%%%%%%%%%%%%% referenc.tex %%%%%%%%%%%%%%%%%%%%%%%%%%%%%%
% sample references
% "physics"
%
% Use this file as a template for your own input.
%
%%%%%%%%%%%%%%%%%%%%%%%% Springer-Verlag %%%%%%%%%%%%%%%%%%%%%%%%%%


% Monographs
%\bibitem{monograph} H. Ibach, H. L\"uth: \textit{Solid-State
%Physics}, 2nd edn (Springer, Berlin Heidelberg New York 1996) pp 45--56

% Contributed Works
%\bibitem{contribution} D.M. MacKay: Visual stability and voluntary eye
%movements. In: \textit{Handbook of Sensory Physiology}, vol 3, ed by R.
%Jung, D.M. MacKay (Springer, Berlin Heidelberg New York 1973) pp
%307--331

% Journal
%\bibitem{journal} S. Preuss, A. Demchuk Jr, M. Stuke et al: Appl. Phys.
%A \textbf{61}, 33 (1995)

% Theses
%\bibitem{thesis} D.W.  Ross: Lysosomes and storage diseases. MA
%Thesis, Columbia University, New York (1977)

%\end{thebibliography}

%%%%%%%%%%%%%%%%%%%%%%%%%%%%%%%%%%%%%%%%%%%%%%%%%%%%%%%%%%%%%%%%%%%%%%  

%%%%%%%%%%%%%%%%%%%%%%%%%%%%%%%%%%%%%%%%%%%%%%%%%%%%%%%%%%%%%%%%%%%%%%

\printindex

\begin{thebibliography}{99.}
% Use the following syntax and markup for your references
%

\bibitem{asympfree} D.~J.~Gross and F.~Wilczek: Phys. Rev. Lett.  {\bf 30}, 1343 (1973);\\
                    H.~D.~Politzer: Phys. Rev. Lett.  {\bf 30}, 1346 (1973)

\bibitem{lattice} I.~Montvay and G.~M\"unster, ``Quantum fields on a lattice,''
                  Cambridge University Press, Cambridge (1994);\\
                  R.~Gupta: [arXiv:hep-lat/9807028]                  
                  
                  
\bibitem{Wein1} S.~Weinberg: Physica {\bf 96} A, 327 (1979)

\bibitem{GL1} J.~Gasser and H.~Leutwyler: Ann. Phys. {\bf 158}, 142 (1984)

\bibitem{GL2} J.~Gasser and H.~Leutwyler: Nucl. Phys. B {\bf 250}, 465 (1985)

\bibitem{ChPT}  See e.g:\\
                V.~Bernard, N.~Kaiser and U.-G.~Mei{\ss}ner: Int. J. Mod. Phys. E {\bf 4}, 193 (1995);\\
                V.~Bernard and U.-G.~Mei{\ss}ner: [arXiv:hep-ph/0611231];\\
                G.~Ecker: Prog. Part. Nucl. Phys.  {\bf 35}, 1 (1995);\\
                G.~Ecker: [arXiv:hep-ph/0011026];\\
                J.~Gasser: Lect. Notes Phys.  {\bf 629}, 1 (2004);\\
                H.~Leutwyler: [arXiv:hep-ph/9406283];\\
                H.~Leutwyler: [arXiv:hep-ph/0008124];\\
                A. V. Manohar: [arXiv: hep-ph/9606222];\\
                U.-G. Mei{\ss}ner:  Rept. Prog. Phys.  {\bf 56}, 903 (1993);\\
                A.~Pich: [arXiv:hep-ph/9806303];\\
                S.~Scherer: Adv. Nucl. Phys.  {\bf 27}, 277 (2003);\\
                S.~Scherer and M.~R.~Schindler: [arXiv:hep-ph/0505265]

\bibitem{HE} W.~Heisenberg and H.~Euler: Z. Physik \textbf{98}, 714 (1936)

\bibitem{Schw} J.~Schwinger: Phys. Rev. \textbf{82}, 664 (1951)

\bibitem{pdg} W.-M.~Yao {\it et al.} [Particle Data Group Collaboration]:
              %``Review of particle physics,''
              J.\ Phys.\ G {\bf 33}, 1 (2006) 
              
\bibitem{NJL} Y.~Nambu: Phys. Rev. Lett. {\bf 4}, 380 (1960); Phys. Rev. {\bf 117}, 648 (1960);\\
              Y.~Nambu and G.~Jona-Lasinio: Phys. Rev. {\bf 122}, 345 (1961) 
              
\bibitem{Hol} B.~R.~Holstein: Phys. Lett. B {\bf 244}, 83 (1990) 

\bibitem{GenChPT} N.~H.~Fuchs, H.~Sazdjian and J.~Stern:  Phys. Lett. B {\bf 269}, 183 (1991);\\
                  M.~Knecht, H.~Sazdjian, J.~Stern and N.~H.~Fuchs:  Phys. Lett. B {\bf 313}, 229 (1993)

\bibitem{SVZ} M.~A.~Shifman, A.~I.~Vainshtein and V.~I.~Zakharov: 
              Nucl. Phys. B {\bf 147}, 385 (1979)
              
\bibitem{Dash} R.~Dashen: Phys. Rev. {\bf 183}, 1245 (1969)                   
              
\bibitem{Wei1} S.~Weinberg: Trans. NY Acad. Sci. {\bf 38}, 185 (1977)

\bibitem{BCE2} J.~Bijnens, G.~Colangelo and G.~Ecker: JHEP {\bf 9902}, 020 (1999)

\bibitem{GJLW} I.~Gerstein, R.~Jackiw, B.~W.Lee and S~Weinberg: Phys. Rev. D {\bf 3}, 2486 (1971) 

\bibitem{BCE} J.~Bijnens, G.~Colangelo and G.~Ecker: Annals Phys. {\bf 280}, 100 (2000)

\bibitem{LEClat} See, e.g., C.~Aubin {\it et al.}  [MILC Collaboration]:
                 Phys. Rev. D {\bf 70}, 114501 (2004)

\bibitem{BEG} J.~Bijnens, G.~Ecker and J.~Gasser: [arXiv:hep-ph/9411232]

\bibitem{reso} G.~Ecker, J.~Gasser, A.~Pich and E.~de Rafael: Nucl. Phys. B {\bf 321}, 311 (1989);\\
               J.~F.~Donoghue, C.~Ramirez and G.~Valencia: Phys. Rev. D {\bf 39}, 1947 (1989)   

\bibitem{Dashcorr} J.~F.~Donoghue, B.~R.~Holstein and D.~Wyler: Phys. Rev. D {\bf 47}, 2089 (1993);\\
               R.~Baur and R.~Urech: Phys. Rev. D {\bf 53}, 6552 (1996);\\
               J.~Bijnens and J.~Prades: Nucl. Phys. B {\bf 490}, 239 (1997)

\bibitem{Qval} J.~Kambor, C.~Wiesendanger and D.~Wyler: Nucl. Phys. B {\bf 465}, 215 (1996);\\
               A.~V.~Anisovich and H.~Leutwyler: Phys. Lett. B {\bf 375}, 335 (1996);\\
               B.~V.~Martemyanov and V.~S.~Sopov: Phys. Rev. D {\bf 71}, 017501 (2005)

\bibitem{anomalies} W.~A.~Bardeen: Phys. Rev. {\bf 184}, 1848 (1969);\\
                    K.~Fujikawa: Phys. Rev. Lett. {\bf 42}, 1195 (1979)

\bibitem{Wein} S.~Weinberg: Phys. Rev. D {\bf 11}, 3583 (1975)

\bibitem{etap} E.~Witten: Ann. Phys. (N.Y.) {\bf 128}, 363 (1980);\\
                  P.~di Vecchia and G.~Veneziano: Nucl. Phys. B {\bf 171}, 253 (1980)

\bibitem{largeNc} R.~Kaiser and H.~Leutwyler: Eur. Phys. J. C {\bf 17}, 623 (2000)

\bibitem{EDM} A.~Pich and E.~de Rafael: Nucl. Phys. B {\bf 367}, 313 (1991);\\
              B.~Borasoy: Phys. Rev. D {\bf 61}, 114017 (2000)




\end{thebibliography}
\end{document}